\documentclass[12pt]{iopart}
\usepackage{graphicx}
\usepackage{xcolor}
\usepackage{tikz}
\usepackage{stackrel}
\usepackage{dsfont}
\usepackage{bm}
\usepackage{xcolor,cancel}
\usepackage{ulem}
\usepackage{hyperref}
\usepackage{pifont}

\newcommand*\circled[1]{\tikz[baseline=(char.base)]{
            \node[shape=circle,draw,inner sep=1pt] (char) {#1};}}

\begin{document}
\title[Two-Mode Mechanical Entanglement Through Pulsed Optomechanical Measurements]{Preparation and Verification of Two-Mode Mechanical Entanglement Through Pulsed Optomechanical Measurements}

\author{P. Neveu$^{1}$, J. Clarke$^{2}$, M. R. Vanner$^{2}$, E. Verhagen$^{1}$}

\address{$^{1}$Center for Nanophotonics, AMOLF, Science Park 104, 1098 XG Amsterdam, The Netherlands}
\address{$^{2}$QOLS, Blackett Laboratory, Imperial College London, London SW7 2BW, United Kingdom}

\begin{abstract}
We propose a protocol how to generate and verify bipartite Gaussian entanglement between two mechanical modes coupled to a single optical cavity, by means of short optical pulses and measurement. Our protocol requires neither the resolved sideband regime, nor low thermal phonon occupancy, and allows the generation and verification of quantum entanglement in less than a mechanical period of motion. Entanglement is generated via effective two-mode mechanical squeezing through conditioning position measurements. We study the robustness of entanglement to experimental deviations in mechanical frequencies and optomechanical coupling rates.
\end{abstract}

\submitto{\NJP}
\maketitle

\section{Introduction}
Quantum entanglement is a fundamental resource of quantum technologies \cite{horodecki2009quantum}. Besides the intriguing concept of correlations much stronger than classically understandable \cite{PhysRev.47.777}, the property of inseparability of systems at the quantum level has been widely exploited \cite{erhard2020advances} towards the implementation of quantum teleportation \cite{pirandola2015advances}, computing \cite{preskill2018quantum}, communication \cite{RevModPhys.74.145,ursin2007entanglement}, fundamental tests of quantum mechanics \cite{hensen2015loophole, PhysRevLett.115.250401, PhysRevLett.115.250402}, entanglement-enhanced measurements \cite{guo2020distributed,zhuang2018distributed} or more generally quantum sensors \cite{RevModPhys.89.035002}. Entanglement of the motional degree of freedom is now being actively study on the theoretical side \cite{PhysRevLett.88.120401, Pinard_2005, Li_2015}, with proposals based on reservoir engineering \cite{chauhan2020stationary, PhysRevLett.98.030405, borkje2011proposal, PhysRevA.87.033829}, conditioning measurement \cite{clarke2020generating, brunelli2020stroboscopic,Bennett_2016,Vostrosablin_2018} and swapping of optical entanglement \cite{pirandola2006macroscopic, zhang2003quantum,vostrosablin2016pulsed,vacanti2008optomechanical}. Experimental implementations have been tested in systems ranging from trapped ions \cite{jost2009entangled}, optical phonons in diamond \cite{lee2011entangling} to photonic crystals \cite{riedinger2018remote} and microwave eletromechanical resonators \cite{ockeloen2018stabilized}.

So far, optomechanically-induced mechanical entanglement has been achieved in systems that require the mechanical frequency $\omega$ to be much larger than a pulse or cavity bandwidth $\kappa$  \cite{lee2011entangling,riedinger2018remote, ockeloen2018stabilized}. This resolved sideband regime can be challenging to achieve as it requires a quality factor $Q_\mathrm{em}$ of the electromagnetic mode to be higher than the ratio of electromagnetic frequency $\omega_\mathrm{em}$ over mechanical frequency $\omega$. For systems that do not resolve motional sidebands ($\kappa\gg\omega$), an alternative can be sought through conditioning the mechanical state with snapshot measurements of the mechanical position, using optical pulses shorter than the mechanical period \cite{vanner2011pulsed}. Provided that the mechanical-position-induced phase shift can be well resolved above the optical shot noise, significant mechanical cooling or squeezing by measurement can be achieved \cite{vanner2013cooling}. Indeed, using this technique, mechanical squeezing approaching the quantum regime has been achieved, thus providing a promising route for mechanical entanglement \cite{muhonen2019state,clarke2020generating,brunelli2020stroboscopic}.

In \cite{clarke2020generating}, a protocol is proposed for mechanical entanglement and verification of two spatially-separated optomechanical resonators, which can be essential for entanglement distribution applications. Routes for generation and verification of entanglement by combining pulsed measurements on the separated systems are then identified. Reference \cite{brunelli2020stroboscopic} proposes to create two-mode entanglement of a single device using measurements in the so-called \textit{stroboscopic} limit, with a protocol that does not explicitly depend on the mechanical mode frequencies. The sensitivity of the expected degree of entanglement to experimental deviations such as optical matching of spatially-separated devices, or optomechanical coupling mismatch between the mechanical modes, remains unexplored.

In this work, we explore the ability to generate and verify entanglement of two mechanical modes of a single optomechanical system, in a minimal two-pulse measurement sequence, lasting for less than a mechanical period of both resonators. As such, our methods differs from those using measurements on separate cavities \cite{clarke2020generating} or using stroboscopic measurements \cite{brunelli2020stroboscopic} of a single device. These two mechanical modes can be two mechanical normal modes of a single body, or separately engineered, for instance with two membranes in a  Fabry-P\'erot cavity \cite{shkarin2014optically,piergentili2018two, gartner2018integrated}, or a sliced photonic crystal nanobeam with two flexural modes of different frequencies \cite{mathew2020synthetic} as sketched in Fig.\,\ref{Protocol}a. We show in section \ref{section2} that the requirements to generate entanglement is as demanding as a pulsed-measurement induced single mode ground-state cooling would require, in terms of optomechanical features, which is within reach with state of the art optomechanical systems \cite{muhonen2019state}. Section \ref{Imp} investigates the robustness of the protocol to experimental deviations. We find that our protocol necessitates a relative frequency tuning of the resonators at percent-level precision, and optomechanical coupling equality at the ten percent level, which make this entanglement scheme realistic from a practical perspective in various opto- and electromechanical platforms.
 
\section{Entanglement and verification sequence}
\label{section2}
\subsection{Qualitative explanation}
The essence of a conditional pulsed optomechanical measurement at a time $t$, is that one can obtain a snapshot of information about the mechanical position of a resonator $x(t)$ by measuring the outgoing light pulse with which it interacted. Reading out the phase quadrature  $P_L$ of the light pulse allows one to measure mechanical position with shot noise imprecision $x_\mathrm{zpf}/\chi$, where $x_\mathrm{zpf}$ is the zero-point fluctuation standard deviation and $\chi\sim \eta \sqrt{N_p} g/\kappa$ is the measurement strength \cite{vanner2011pulsed}, scaling with the optical linewidth $\kappa$, the vacuum optomechanical coupling $g$ and the number $N_p$ of photons in the pulse interacting with the cavity, i.e. the number of incident photons minus input coupling losses. The efficiency $\eta$ incorporates detection and outcoupling efficiency. For large enough $\chi$, one can achieve backaction-evading measurement of the mechanical position, as light shot noise backaction is imparted into the unmeasured momentum quadrature.

Let us consider now that we have two mechanical modes $b_{j=1,2}$ with frequencies $\omega_j$, quadratures $X_j=(b_j+b_j^\dagger)/\sqrt 2$ and $P_j=(b_j-b_j^\dagger)/\sqrt 2 \mathrm i$, and positions $x_j(t)=X_j\cos(\omega_jt)+P_j\sin(\omega_jt)$. Those two modes are assumed for now to interact with the same strength with the optical mode. Then, any conditional pulsed measurement will give information about the total mechanical position $x(t)=x_1(t)+x_2(t)$, leading to a joint measurement of the two resonators at once. This implies strong measurement-induced correlations between the two mechanical modes, and eventually entanglement if the resolution of the collective position is lower than the amplitude of the collective zero-point motion (ZPM). Namely, entanglement of the two-mechanical modes will be achieved if variances of the collective mechanical quadratures are such as
 \begin{equation}
     \Delta X_{+},\Delta P_{-}<1/2\qquad\mathrm{and}\qquad\Delta X_{-},\Delta P_{+}>1/2,
     \label{TMS}
 \end{equation}
 where the collective mode quadrature basis $\left(X_+,P_+,X_-,P_-\right)$ links to the canonical mode quadrature basis $\left(X_1,P_1,X_2,P_2\right)$ by a beamsplitter-like mixing $X_\pm=\left(X_1\pm X_2\right)/\sqrt 2$ and $ P_\pm=\left(P_1\pm P_2\right)/\sqrt 2$. The measured mechanical position
 \begin{equation}
     x(t)=X_1\cos(\omega_1t)+P_1\sin(\omega_1t)+X_2\cos(\omega_2t)+P_2\sin(\omega_2t)
     \label{scan}
 \end{equation}
scans over time multiple collective quadratures because of the frequency mismatch of the two resonators $\omega_1\neq\omega_2$. Then, a measurement of the mechanical position at an arbitrary time $t$ reduces the uncertainty of an arbitrary collective quadrature $X(t)=x(t)/\sqrt 2$ given by Eq.\,(\ref{scan}). We label $b(t)$ the time-dependent collective mode whose amplitude quadrature is $X(t)$, and $c(t)$ to complete the two-mode basis.

\begin{figure}
\begin{centering}
\includegraphics[width=\columnwidth]{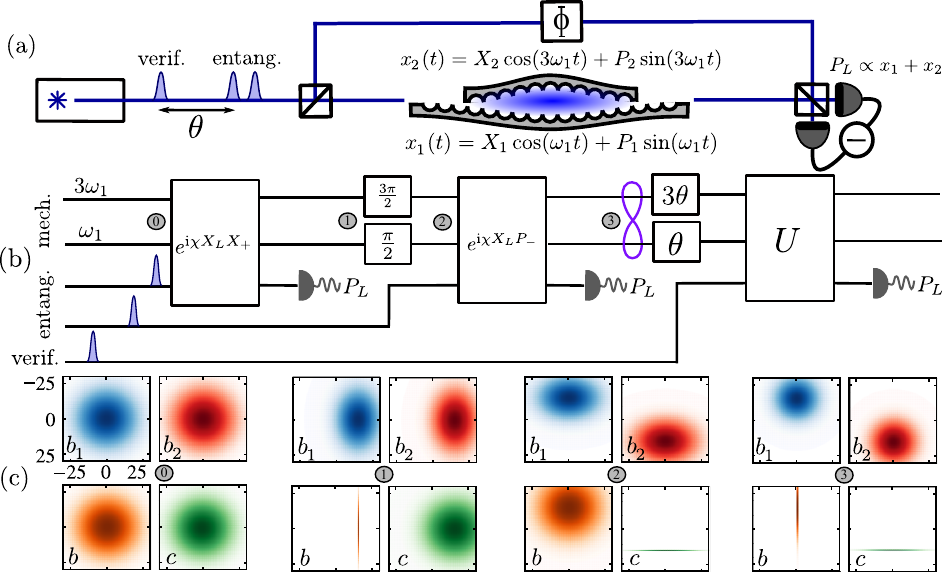}
\par\end{centering}
\caption{\textbf{(a)} Optical pulses shorter than the mechanical period are sent to an optomechanical system. Copies of these pulses are derived and play the role of a local oscillator in the homodyne detection of the outgoing pulses. Adjusting the relative phase $\Phi$ between the two optical paths selects the optical phase quadrature $P_L$ detection. The two mechanical modes can be the two flexural modes of a sliced nanobeam photonic crystal cavity, two octaves detuned. \textbf{(b)} Circuit diagram of the mechanical entanglement and verification protocol. Two pulsed optomechanical measurements of the mechanical position delayed by $\pi/2\omega_1$ allows to get information and then reduces the thermal uncertainty of the $X_+$ and $P_-$ collective quadrature respectively, leading to entanglement of the two resonators in less than a mechanical period. The verification protocol involves an extra pulsed measurement at various delays to reconstruct the covariance matrix of the two-mechanical-modes system. \textbf{(c)} Wigner functions of the partially traced two-mode system over the canonical mode basis ($b_1$, $b_2$) and the collective eigenmode basis ($b$, $c$). Each plot corresponds to different steps of the entanglement protocol.  We assume arbitrary outcome values of $\left<X_+\right> = 15$ and $\left<P_-\right> = 0$ and $n_\mathrm{th}=500$. Measurements affect the collective mechanical quadrature statistics, without significantly affecting the mechanical modes taken individually.\label{Protocol}}
\end{figure}

Starting from a measurement of the position at the time origin $x(0)=\sqrt{2}X_+$, it is not always possible to measure the quadrature $P_-$, which is necessary to fulfill Eq.\,(\ref{TMS}). Indeed, the closer the mechanical frequencies, the longer we need to wait between measurements to measure significantly different collective quadratures, leading to a potential for decoherence to negatively influence the conditional state purity. On the other hand, too much different frequencies can hardly be designed to equally couple to the same optical mode. Then, achieving entanglement shall necessitate multiple measurements to compensate for these drawbacks \cite{brunelli2020stroboscopic}. In this work, we focus on a minimal two-pulse sequence that allows to generate entanglement, independently of the initial thermal occupation of the resonators. For instance, the situation were $\omega_2/\omega_1=3$ allows to entangle and verify in less than a mechanical period of both resonators, while ensuring frequency in the same order of magnitude. This choice of mechanical frequency ratio leads to the two-pulse entanglement  protocol  schemed  in  Fig.\,\ref{Protocol}b. Starting from uncorrelated mechanical resonators (step \circled{0}), the protocol consists of a measurement at the time origin $x(0)=\sqrt{2}X_+$ (step \circled{1}), followed by waiting a delay of a quarter mechanical period of the lowest frequency resonator $\tau = \pi/2\omega_1$ (step \circled{2}) that leads to a mechanical position of $x(\tau)=\sqrt{2}P_-$. A second conditioning measurement at this point (step \circled{3}) generates two-mode squeezing of the mechanical resonators, for large enough measurement strength. 

Figure \ref{Protocol}c shows  Wigner functions of the partially-traced  mechanical  states  in  the canonical ($b_1$, $b_2$) and collective ($b(t)$, $c(t)$) mode basis, conditioned on the measurements, at the different steps of the  entangling sequence. Starting from an initial wide thermal distribution (\circled{0}), uncertainty of the amplitude quadrature of mode $b(0)$ is strongly reduced by the first conditioning measurement (\circled{1}). The two canonical mechanical modes $b_{1,2}$ now have some correlation, which we strengthen with the second entangling pulse. Right before the second entangling pulse (\circled{2}), the mechanical resonators are out of phase, at opposite velocity. At this point, the mechanical collective mode $b(\pi/2\omega_1)$ has the statistics of the collective mode that was unaffected by the first conditional measurement, and its amplitude quadrature is $X(\pi/2\omega_1)=P_-$. After this second conditional measurement (\circled{3}), the mechanical modes are two-mode squeezed. These successive measurements always purify and classically entangle the mechanical resonators, and quantum two-mode squeezing can be achieved if the measurement strength is large enough.

\label{ideal}
\subsection{Quantitative description}

We introduce $\mathbf{X}_\mathrm{can} = \left(X_1,P_1,X_2,P_2,X_L,P_L\right)^\mathrm T$ and $ \mathbf X_\mathrm{col}=\left(X_+,P_+,X_-,P_-,X_L,P_L\right)^\mathrm T$ the quadrature vectors of this bosonic three-mode system in the canonical and collective mode basis, respectively. We note that all the quadratures are defined in the frame rotating at the frequency of their mode, such that their free dynamics is time-independent. The linearized Hamiltonian in the frame rotating at the cavity frequency reads \cite{aspelmeyer2014cavity}
\begin{equation}
    H(t)=\frac{\hbar g}{\sqrt{2}} X_L\left(t\right)\sum_{j}\left(X_j\cos{\left(\omega_jt\right)}+P_j\sin\left(\omega_jt\right)\right),
    \label{Hamiltonian}
\end{equation}
where we assume, for now, that the linearized optomechanical coupling rates of the two mechanical modes are equal to $g$. In the case of a short optical pulse containing $N_p$ photons over a duration $\tau_p\ll1/\omega_{1,2}$ at $t=0$, the evolution operator right after the pulsed interaction writes
\begin{equation}
    U\left(t=0\right)= e^{\mathrm i \chi X_L\left(X_1+X_2\right)/\sqrt{2}},
\end{equation}
where the measurement strength $\chi$ quantifies  the amount of information about mechanical position one can extract from an optical phase quadrature $P_L$ measurement
\begin{equation}
    U\left(0\right) \mathbf X_\mathrm{col} U^\dagger\left(0\right) = \mathbf X_\mathrm{col} + \chi \left(0,X_L,0,0,0,X_+ \right)^{\mathrm{T}}.\end{equation}
Note that we omitted in the evolution operator the deterministic kick on conjugated mechanical momentum $P_1+P_2$ induced by radiation pressure of the mean optical field \cite{vanner2011pulsed}, as it only features a classical displacement. We note that the other orthogonal collective mode ($X_-$, $P_-$) is not affected by this measurement.

Measuring the optical phase quadrature at $t=0$ reduces uncertainty of the $X_+$ quadrature defined previously, while adding backaction noise on $P_+$. More generally, a pulsed measurement at any arbitrary time $t$ always results in the measurement of a combination of quadratures of the two mechanical resonators. We define the time-dependent eigenmode quadrature basis of the Hamiltonian $\mathbf{X}_{\mathrm{eig}}=R_{\omega}(t)R_{\Omega }(t)\mathbf{X}_{\mathrm{col}}=\left(X(t),P(t),Y(t),Q(t),X_L(t),P_L(t)\right)^\mathrm T$ where
\begin{equation}
R_{\omega}\left(t\right)=\left(\begin{array}{cc}
\cos\left(\omega t\right) & \sin\left(\omega t\right)\\
-\sin\left(\omega t\right) & \cos\left(\omega t\right)
\end{array}\right){\oplus}\left(\begin{array}{cc}
\cos\left(\omega t\right) & \sin\left(\omega t\right)\\
-\sin\left(\omega t\right) & \cos\left(\omega t\right)
\end{array}\right)\oplus \mathds{1}_2,
\end{equation}
\begin{equation}
R_{\Omega }\left(t\right)=\left(\begin{array}{cccc}
\cos\left(\Omega t\right) & 0 & 0 & \sin\left(\Omega t\right)\\
0 & \cos\left(\Omega t\right) & -\sin\left(\Omega t\right) & 0\\
0 & \sin\left(\Omega t\right) & \cos\left(\Omega t\right) & 0\\
-\sin\left(\Omega t\right) & 0 & 0 & \cos\left(\Omega t\right)
\end{array}\right)\oplus \mathds{1}_2
\end{equation}
are unitary rotation matrices in the mechanical basis, at the average mechanical frequency $\omega=\left(\omega_1+\omega_2\right)/2$ and relative mechanical frequency $\Omega =\left(\omega_1-\omega_2\right)/2$. We denote by $b(t)$ ($c(t)$) the mechanical collective mode  with quadratures $X(t)$ and $P(t)$ ($Y(t)$ and $Q(t)$), explicitly affected (unaffected) by the interaction. Hamiltonian (\ref{Hamiltonian}) writes $H(t)=\hbar g X_L (t) X(t)$ such that any pulsed measurement of the optical phase quadrature at a time $t$ gives information about $X(t)$ while adding backaction on $P(t)$, and leave the other collective mechanical mode unchanged:

\begin{equation}
    U(t) \mathbf X_\mathrm{eig}U^\dagger\left(t\right) = \mathbf X_\mathrm{eig} + \chi \left(0,X_L,0,0,0,X(t) \right)^{\mathrm{T}}.\end{equation}

The pulsed interaction generates strong correlations between the outgoing light and the mechanical modes. The covariance matrix $\bm{\sigma}=\frac{1}{2}\left<\{\mathbf{(X-\left<X\right>)}, \mathbf{(X-\left<X\right>}) ^\mathrm T\}\right>$ of the three-mode system reads
\begin{equation}
    S_{U(t)} \bm{\sigma} S_{U(t)}^\mathrm{T} = \bm{\sigma}+\chi^{2}\mathrm{diag}\left(0,\frac{1}{2},0,0,0,\sigma_X\right)+\left(\begin{array}{cc}
\mathbf{0} & C\\
C^{\mathrm{T}} & \mathbf{0}
\end{array}\right),
\end{equation}
where $S_{U(t)}$ is the symplectic transformation associated to the operator $U(t)$ \cite{serafini2017quantum,olivares2012quantum}. The first additive term is backaction noise and the second is measurement-induced optomechanical cross-correlations : 
\begin{equation}
    C^{\mathrm{T}}=\chi\left(\begin{array}{cccc}
0 & \frac{1}{2} & 0 & 0\\
\sigma_{X} & \sigma_{XP} & \sigma_{XY} & \sigma_{XQ}
\end{array}\right).
\end{equation}

When measuring the optical phase of the output light, we get information about the mechanical motion. This measurement $\mathcal{M}$ changes our knowledge about the mechanical modes, via the transformation \cite{serafini2017quantum}
\begin{equation}
    \mathcal{M}\left(\bm \sigma_m\right)=\bm \sigma_m - \bm \sigma_{mL}\frac{1}{\bm\sigma_L + \bm\sigma_{\mathrm{HD}}}\bm\sigma_{Lm},\quad\mathrm{with}\quad\bm{\sigma}=\left(\begin{array}{cc}
\bm{\sigma}_m & \bm{\sigma}_{mL}\\
\bm{\sigma}_{Lm} & \bm{\sigma}_{L}
\end{array}\right),
\end{equation}
where $\bm\sigma_{\mathrm{HD}}$ describes the homodyne measurement and $\bm{\sigma}$ is decomposed over the mechanical ($\bm{\sigma}_m$) and optical ($\bm{\sigma}_L$) basis, and their cross-correlations ($\bm{\sigma}_{mL}$). If we consider to have an ideal non-lossy homodyne detection locked on the phase quadrature of the outgoing light field, we have \cite{serafini2017quantum} 

\begin{equation*}
   ( \bm\sigma_{L}+\bm\sigma_{\mathrm{HD}})^{-1}\rightarrow\mathrm{diag}\left(0,1/\sigma_{L}^{22}\right)=\mathrm{diag}\left(0,\left(\frac{1}{2}+\chi^2 \sigma_X\right)^{-1}\right).
\end{equation*}
Finally, the effect of a single measurement at time $t$ can be summed up in the mechanical covariance matrix in the canonical basis as $
    \bm\sigma_m' = P(t) \mathcal{M}\left(P(t)^{-1} \bm\sigma_m P(t)\right)P(t)^{-1} $,
where $P(t)=R_{\omega}\left(t\right)R_{\Omega }\left(t\right)$. Then, the entangling sequence of Fig.\,\ref{Protocol}b leads to the exact following covariance matrix in the canonical mode basis:
\begin{equation}
    \bm\sigma_m^\mathrm{ent}=\left(\begin{array}{cc}
A_{1} & C_{12}\\
C_{12}^{\mathrm{T}} & A_{2}
\end{array}\right),\,\mathrm{with\,}\begin{array}{l}
A_{1}=\left(n_{1}+\frac{\chi^{2}}{2}-\frac{2n_{1}^{2}\chi^{2}}{1+2\left(n_{1}+n_{2}\right)\chi^{2}}\right)\mathds{1}_{2}\\
A_{2}=A_{1}+\frac{n_{2}-n_{1}}{1+2\left(n_{1}+n_{2}\right)\chi^{2}}\mathds{1}_{2}\\
C_{12}=\left(\frac{\chi^{2}}{2}+\frac{2n_{1}n_{2}\chi^{2}}{1+2\left(n_{1}+n_{2}\right)\chi^{2}}\right)\sigma_z,
\end{array}
\end{equation}
where $n_i = \frac{1}{2}+ n_{\mathrm{th},i}$ is the total occupation of the mechanical mode $i$, and $\sigma_z$ the $z-$Pauli matrix. The asymmetry of this covariance matrix between the two mechanical modes ($A_1 \neq A_2$) originates from the initial unequal thermal occupation of the two modes ($n_{1}\neq n_{2}$). The covariance matrix in the canonical basis can be explicitly written as a two-mode squeezed state by parameterizing its eigen-decomposition as $\mathrm{diag}\left(n_e e^{-r_e},n_e e^{r_e},n_e e^{r_e},n_e e^{-r_e}\right)$ with $n_e e^{\pm r_e} = \left(\alpha_{+}+\beta\pm\sqrt{\alpha_{-}^{2}+\beta^{2}}\right)/2$,
\begin{equation}
    \alpha_\pm = \frac{n_{1}\pm n_{2}}{1+2\left(n_{1}+n_{2}\right)\chi^{2}}\quad\mathrm{and}\quad\beta =\chi^2\left(1+ \frac{4n_{1}n_{2}}{1+2\left(n_{1}+n_{2}\right)\chi^{2}}\right).
\end{equation}
In the relevant regime of initial high thermal occupations $n_{i}\gg \{1,\chi\}$ these eigenvalues simplify to
\begin{equation}
     n_e e^{- r_e} = \frac{1}{4\chi^2}\qquad\mathrm{and}\qquad n_e e^{+ r_e} = \frac{2n_{1}n_{2}}{n_{1}+n_{2}}.
     \label{eigenval}
\end{equation}
The first term corresponds to the measurement resolution associated with the $X_+$ and $P_-$ collective quadratures measurement. The second term corresponds to the standard deviation of the unmeasured collective quadratures, with thermal occupation averaged over the two mechanical modes.

The duration of the pulse is considered to be negligible compared to the decoherence decay rate. However, decoherence of the mechanical modes is taking place in between measurement pulses, i.e. during the free evolution of the mechanical modes. We take it into account by considering the mixing of the prepared state with a thermal state at rate $\Gamma_{1,2}$:
\begin{equation}
    {\mathcal{D}}\left(\bm\sigma_m,t\right)=\sqrt{\mathds{G}}\bm\sigma_m\sqrt{\mathds{G}}+\left(\mathds{1}_4-\mathds{G}\right)\mathrm{diag}\left(n_\mathrm{1},n_\mathrm{1},n_\mathrm{2},n_\mathrm{2}\right).
    \label{Decoh}
\end{equation}
with $\mathds{G}=\mathrm{diag}\left(e^{-\Gamma_1 t},e^{-\Gamma_1 t},e^{-\Gamma_2 t},e^{-\Gamma_2 t}\right)$ and $\Gamma_in_i$ the heating rate of mechanical mode $i$ \cite{ferraro2005gaussian}. For sake of clarity, we focus on the relevant regime of high--quality thermal resonators, i.e. with mechanical quality factors $Q_i=\omega_i/\Gamma_i\gg1$ and $n_{1}=3n_{2}=n\gg1$. In this regime, the full entangling sequence of Fig.\,\ref{Protocol}b leads to the following covariance matrix eigen-decomposition :
\begin{equation}
    \bm{\sigma}_m^\mathrm{ent} = \mathrm{diag}\left(\frac{1}{4\chi^2}+\frac{\pi n}{4}\left(\frac{1}{Q_1}+\frac{1}{Q_2}\right),\frac{n}{2},\frac{n}{2},\frac{1}{4\chi^2}\right)+\mathcal{O}\left(\frac{1}{Q_i^2}\right).
    \label{covapprox}
\end{equation}
where we have developed at the leading order in $1/Q_i$. This covariance matrix can be parametrized with the relevant two-mode squeezing parameters $\mathrm{diag}\left(n_+ e^{-r_+},n_+ e^{r_+},n_- e^{r_-},n_- e^{-r_-}\right)$. The collective quadratures $X_+$ and $P_-$ ($n_{\pm}e^{-r_\pm}$ terms) are squeezed below the ZPM amplitude, provided that the measurement strength exceeds unity and that decoherence does not induce noise on the $X_+$ quadrature larger than the ZPM amplitude in between the two pulses. The two other collective quadratures ($n_{\pm}e^{+r_\pm}$ terms) are only affected by measurement quantum backaction, which is usually very weak with respect to the wide thermal variance. We note that $P_-$ is in principle not affected at all by decoherence because the covariance matrix above is written right after the second measurement.

We describe the amount of entanglement by means of the logarithmic negativity $E_N$. This quantity estimates the maximal number of singlet Bell states one could get with an entanglement distillation protocol. Besides the advantage of being measurable \cite{vidal2002computable}, this quantity is widely considered in both discrete \cite{peres1996separability,audenaert2003entanglement} and continuous variable \cite{simon2000peres} entanglement characterization. It characterises the deviation from positiveness of the two-mode system after a single-mode transposition has been applied. After transforming the covariance matrix (\ref{covapprox}) back to the  canonical basis, an analytical derivation of the logarithmic negativity of a bipartite Gaussian system (\ref{covapprox}) leads to \cite{vidal2002computable}
\begin{equation}
    E_\mathrm{N}^\mathrm{ent} = \mathrm{max}\left(0,-\mathrm{log}_2\left(2 \sqrt{ n_-n_+ e^{-r_--r_+}}\right)\right).
    \label{Entwomode}
\end{equation}
Then, entanglement can be efficiently generated as soon as the geometric mean of the two squeezed quadrature variances is lower than the ZPM amplitude.

From now on, for sake of simplicity, we suppose the mechanical quality factors of both resonators to be equal $Q_1=Q_2=Q$, meaning that the thermal decoherence rates $\Gamma_in_{i}$ are equal too. Figure \,\ref{EstColLogNeg}a shows the logarithmic negativity as a function of the mechanical quality factor, for different measurement strength. In the relevant approximated regime of Eq.\,(\ref{covapprox}), the criteria to achieve generation of entanglement are
\begin{equation}
    E_N^\mathrm{ent} = -\log_2\left(\frac{1}{2\chi^2}\sqrt{1+\frac{2\pi n\chi^2}{Q}}\right)>0\Leftrightarrow\frac{\chi^4-1}{\chi^2}>\frac{2\pi n}{Q}\;\mathrm{and}\;\chi>\frac{1}{\sqrt{2}}.
    \label{CriteriaEnt}
\end{equation}
These requirements are not more demanding than those already required for conditional ground state cooling or squeezing of a single mechanical mode using conditional pulsed measurements \cite{vanner2011pulsed}. Recent experiments with a single flexural mode of a sliced nanobeam photonic crystal structure with moderate mechanical properties ($\omega_m/2\pi\simeq 3\,$MHz, $\kappa/2\pi\simeq20\,$GHz, $g/2\pi \simeq 25\,$MHz, $\eta\simeq1$\,\%, $N_p = 60$, $Q \simeq 3\times10^5$) have already reached a measurement strength of $\chi=0.11$ \cite{muhonen2019state}, highlighting that achieving the criteria above is within reach of near-term experiments. We note that a Brownian model may be used to model the mechanical decoherence, in which the mechanical momentum is displaced through interactions with the bath, instead of the symmetric beamsplitter model described by Eq.\,(\ref{Decoh}). If  one does this, the criteria to generate entanglement will be less stringent than Eq.\,(\ref{CriteriaEnt}) as the optical phase quadrature measurement contains information about mechanical position only, which is unaffected by a Brownian model until free evolution of the mechanical mode evolves the momentum noise into position.

We finally note that two-mode squeezing can in principle be achieved with a single pulse measurement, in case of low thermal occupation. Indeed, if one measures the $X_+$ quadrature only, the conditioned covariance matrix eigen-decomposition writes  $\mathrm{diag}\left(1/4\chi^2,n/2,n/3,n\right)$. One can show that this two-mode squeezed state is quantum entangled if the challenging criteria $E_\mathrm N^\mathrm{ent}>0\Leftrightarrow n<4\chi^2$ is fulfilled.

\subsection{Verification of entanglement}

\begin{figure}
\begin{centering}
\includegraphics[width=\columnwidth]{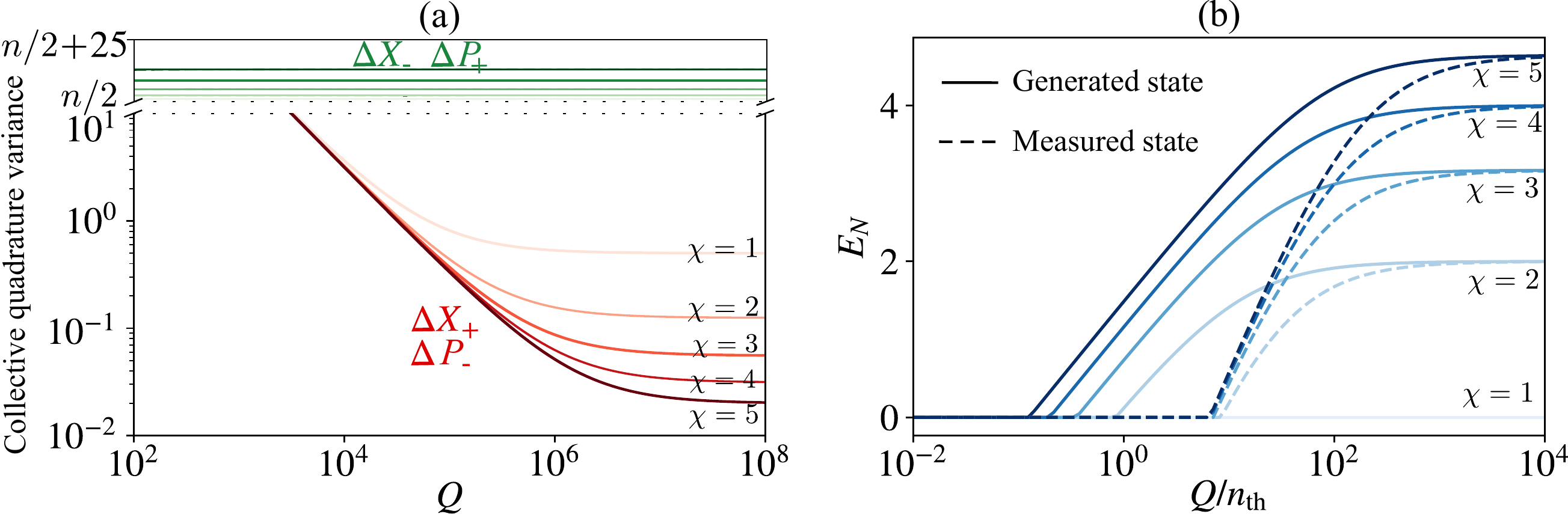}
\par\end{centering}
\caption{(a) Estimated collective variance from the verification protocol  against mechanical quality factor $Q$. Color contrast scales with the measurement strength $\chi$, valued from 1 (shaded color) to 5 (dusky color). Noise variance of the $X_+$ and $P_-$ quadratures can be strongly reduced even below the ZPM amplitude, while the noise variance of the $X_-$ and $P_+$ quadratures are slightly affected by backaction, because of the relatively large thermal phononic occupation $n_\mathrm{th}=10^4\gg\chi^2$. Thermal decoherence between pulses tends to thermalize the squeezed thermal state. (b) Logarithmic negativity of the state generated after entanglement sequence (continuous line) and the evaluated state by verification sequence (dashed line). The verification protocol always underestimates the amount of entanglement present in the two-mode mechanical state, and tends to optimally estimate it in the low-decoherence limit.}
\label{EstColLogNeg}
\end{figure}

A complete verification protocol of the entanglement could require one to reconstruct the full two-mode covariance matrix of the mechanical state. However, we can only access a restricted ensemble of collective quadratures $X\left(t\right)$ by means of mechanical position measurement. Nevertheless, we use the fact that the covariance matrix (\ref{covapprox}) is diagonal in the $\mathbf X_\mathrm{col}$ basis to selectively detect each of its diagonal components \cite{note}.  The $X_{+}$ ($P_{-}$) quadrature variance $\sigma_{X_+}^\mathrm{ent}$ ($\sigma_{P_-}^\mathrm{ent}$) can be measured independently and selectively by waiting a delay $t = \pi/2\omega_1$ ($t = \pi/\omega_1$) from the last entangling pulse. Namely, multiple measurements would distribute over a Gaussian with width $\sigma_{X_+}^\mathrm{ver}= \mathcal{D}\left(\sigma_{X_+}^\mathrm{ent},\pi/2\omega_1\right)$ ($\sigma_{P_-}^\mathrm{ver}= \mathcal{D}\left(\sigma_{P_-}^\mathrm{ent},\pi/\omega_1\right)$). The two other collective quadratures cannot be measured selectively. One option is to measure the variance of the superposition $\sigma_{X_-+P_+}$ by waiting a delay $t = \pi/4\omega_1$ from the last entangling pulse, and assume that the thermal distribution of the superposition is the same as the distribution of $X_-$ and $P_+$ taken individually :  $\sigma_{X_-}^\mathrm{ver}=\sigma_{P_+}^\mathrm{ver}= \mathcal{D}\left(\sigma_{X_-+P_+}^\mathrm{ent},\pi/4\omega_1\right)$. This assumption is exact in the ideal case of resonators detuned by two octaves. We will investigate in the next section to which extent this approximation is valid for entanglement verification for detuned resonators.

The reconstructed mechanical covariance matrix in the canonical mode basis from verification is then written as $\bm{\sigma}_m^\mathrm{ver} = \mathrm{diag}\left(\sigma_{X_+}^\mathrm{ver},\sigma_{X_-}^\mathrm{ver},\sigma_{P_+}^\mathrm{ver},\sigma_{P_-}^\mathrm{ver}\right) $. In the relevant parameter regime of Eq.\,(\ref{covapprox}), it writes
\begin{equation}
    \bm{\sigma}_m^\mathrm{ver} = \mathrm{diag}\left(\frac{1}{4\chi^2}+\frac{\pi n}{Q},\frac{n}{2},\frac{n}{2},\frac{1}{4\chi^2}+\frac{\pi n}{Q}\right).
\end{equation}

Figure \ref{EstColLogNeg}a shows the collective quadrature variances evaluated by verification protocol. Collective quadratures that were conditionally measured during the entangling sequence are drastically reduced below the thermal level. We note that both variances $\Delta X_+$ and $\Delta P_-$ are equally evaluated because of the same time delay -- i.e. decoherence thermal mixing -- between their respective conditional measurement pulse and verification pulse. The logarithmic negativity of this measurable covariance matrix will always underestimate the actual amount of entanglement right after the entangling sequence $E_N^\mathrm{ver}=-\mathrm{log}_2\left(\frac{1}{2\chi^2}+\frac{2\pi n}{Q}\right)<E_N^\mathrm{ent}$. The criteria to measure entanglement are then
\begin{equation}
    \frac{2\pi n}{Q}<1\quad\mathrm{and}\quad\chi>\frac{1}{\sqrt{2}},
\end{equation}
where the first condition can be interpreted as the requirement for the conditional quadrature variance to remain below the ZPM amplitude after half a mechanical period waiting delay, starting from an infinitely squeezed state.

The logarithmic negativity of the evaluated mechanical state together with the entangled state is plotted in Fig.\,\ref{EstColLogNeg}b. The requirement on the mechanical quality factor is more stringent to detect entanglement than to generate it because of mechanical decoherence and rethermalization between the entangling and verification pulses. 

\section{Robustness of the protocol to experimental imperfections}
\label{Imp}

The protocol we propose requires having two mechanical resonators tuned at two octaves from each other, that moreover couple to an optical cavity with the same rate. These requirements can be experimentally challenging if one imagines to entangle two higher order modes of the same mechanical resonator for example. We discuss in this section how stringent these requirements are, by considering deviations of the mechanical frequencies (section \ref{MechImp}) and the coupling strength (section \ref{CoupImp}) from the ideal protocol.

\subsection{Mechanical frequencies imprecision}
\label{MechImp}

\begin{figure}
\begin{centering}
\includegraphics[width=\columnwidth]{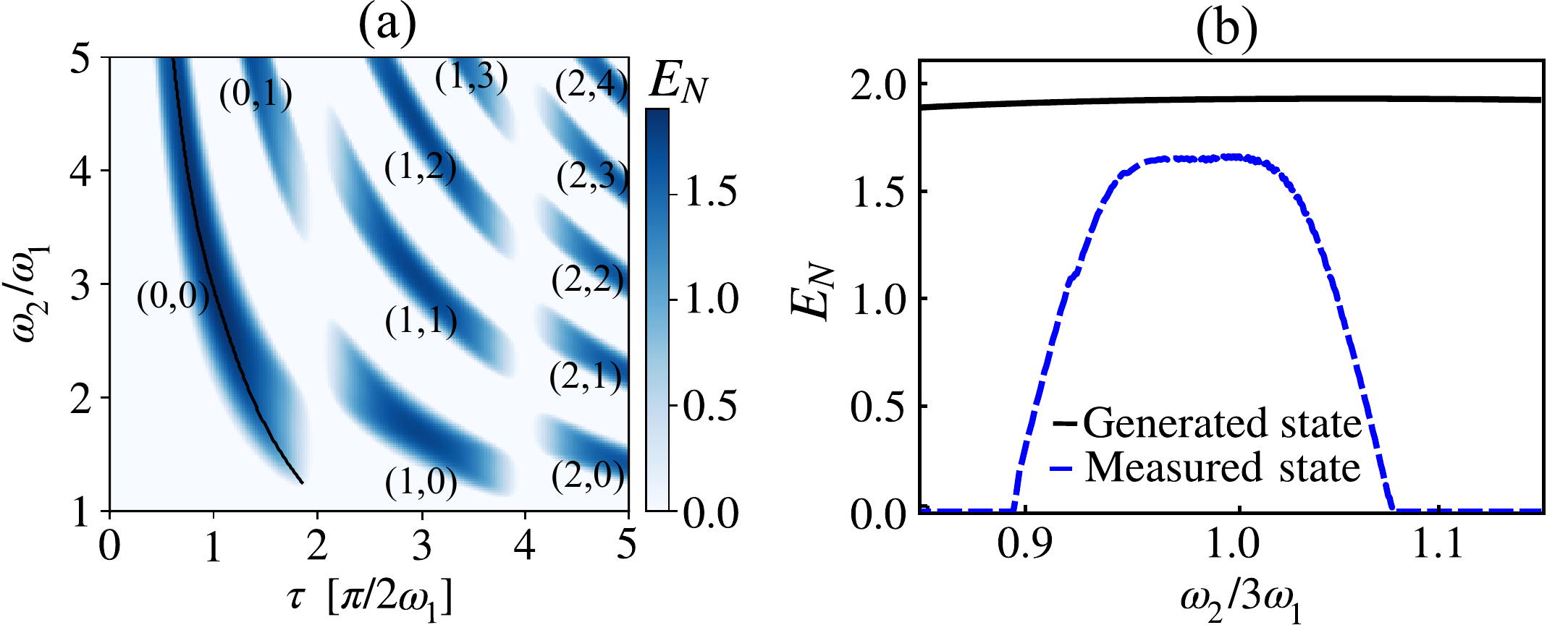}
\par\end{centering}
\caption{Entanglement \textbf{(a)} and verification \textbf{(b)} protocols robustness against mechanical frequencies imprecision. $Q=10^6$, $\chi=2$, $n_\mathrm{th} = 10^4$. \textbf{(a)} Each blue area correspond to two-pulse sequences ($k,l$) leading to mechanical entanglement between the two resonators. The ($0,0$) case is the closest to the origin, leading to the less decoherence losses ($\tau$ is the lowest) and have the experimental advantage of closest mechanical resonator frequencies ($\omega_2/\omega_1$ the lowest). Mechanical frequencies mismatch can be compensated by adjusting the delay $\tau$ between the two pulses of the entanglement sequence (black line). \textbf{(b)} Black line represents the logarithmic negativity of the $\tau$-optimized entanglement protocol (0,0). Dashed line shows the optimized measurable logarithmic negativity, by playing on the three delays of the verification pulses. Maximum measurable $E_N$ does not reach actual value because of decoherence during the verification sequence. \label{FreqImp}}
\end{figure}  

Figure \ref{FreqImp}a shows numerically computed logarithmic negativity of the mechanical state generated, for a wide range of two-pulse entanglement sequence, and over a wide range of mechanical frequencies ratio. Plots are obtained in the case of a modest measurement strength $\chi=2$. Each blue area is centered on an efficient two-pulse entanglement sequence characterized by the doublet $(k,l)\in\mathds{Z}^+$ satisfying :
\begin{equation}
    \tau\omega_1=\left(2k+1\right)\frac{\pi}{2}\qquad\mathrm{and}\qquad\tau\omega_2=\left(2k+3\right)\frac{\pi}{2}+2l\pi.
\end{equation}
These two conditions mean that the second pulse should be applied after a delay $\tau$ after which the two mechanical oscillators are in opposite phase and at maximum velocity. All these protocols are in principle equivalent. However, increasing values of $k$ require longer waiting times, thus causing more impact of thermal decoherence. Moreover, we can see qualitatively that the width of the blue regions in the $\omega_2/\omega_1$ direction decreases, with increasing values of $\omega_2/\omega_1$. We have chosen to focus in the former section on the ($k=0$, $l=0$) protocol, because of its lowered sensitivity to decoherence and the practical advantage of close mechanical frequencies. We moreover observe numerically that the entanglement generated by this protocol is the most resilient to experimental detuning of the resonators. Last but not least, the width of the blue regions indicates that one can compensate for any mechanical frequency detuning by shifting the delay between the two pulses of the entanglement sequence.

Figure \ref{FreqImp}b compares the logarithmic negativity reachable by the entanglement sequence with the one obtained with the verification sequence, in the case of the ($k=0$, $l=0$) protocol. The entanglement sequence shows strong robustness to mechanical detunings, as it can be compensated by adjusting the delay between the entangling pulses, by following the black line of Fig.\,\ref{FreqImp}a. Similarly, entanglement verification can be adapted by adjusting the delays of the verification pulses, such that entanglement can be optimally verified. Figure \ref{FreqImp}b shows that the requirement on the mechanical frequency tuning to verify entanglement is more stringent than for its generation, on the order of $\pm5$\% for a measurement strength of $\chi=2$. Moreover, the verification protocol imprecision requirement is not centered on the ideal case where $\omega_2=3\omega_1$. This asymmetry is because $\omega_2\geq3\omega_1$ protocols are slightly more affected by decoherence because the optimally adjusted pulse delays are longer than $\pi/2\omega_1$. We interpret the fact that verification sequence is much more sensitive to frequency mismatch than the entanglement sequence because of the restricted number of component of the covariance matrix we can actually measure, while assuming all other non-diagonal terms vanish.

\subsection{Optomechanical coupling imprecision}
\label{CoupImp}

\begin{figure}
\begin{centering}
\includegraphics[width=\columnwidth]{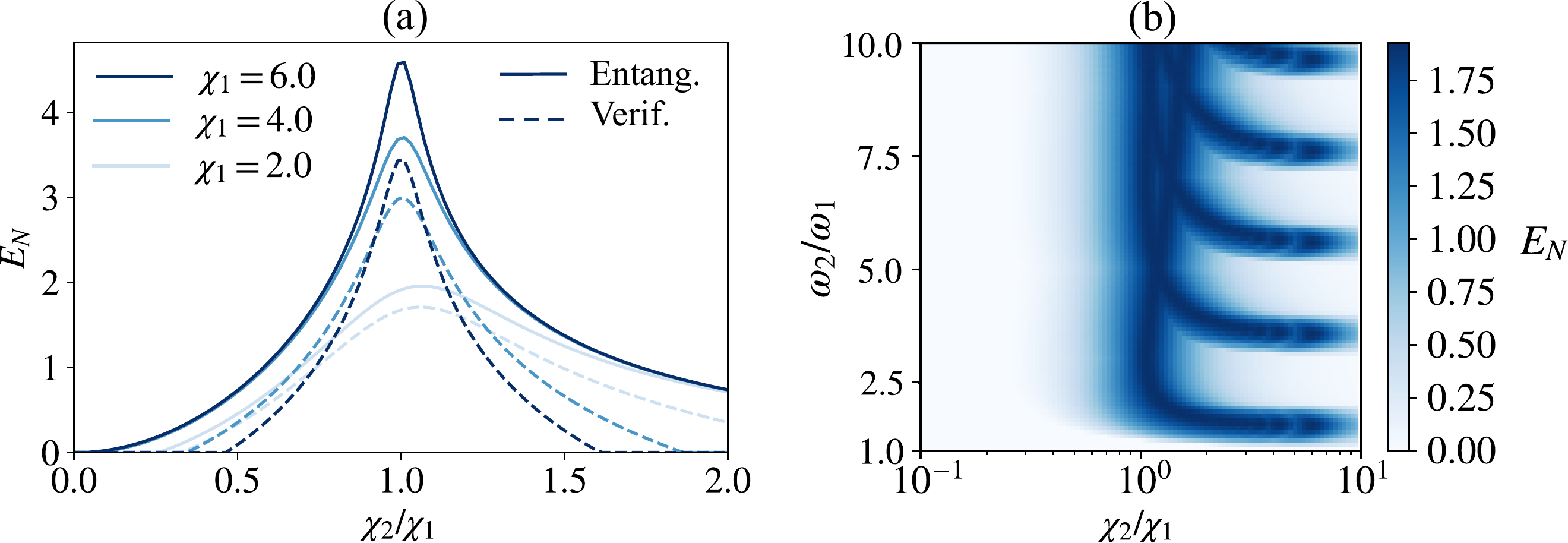}
\par\end{centering}
\caption{\textbf{(a)} Logarithmic negativity of the state generated after the entanglement sequence (continuous) and of the evaluated state with the verification sequence (dashed) versus the measurement strength mismatch of the two resonators. Various measurement strengths of the lowest frequency resonator are considered. Measured entanglement by the verification sequence is reduced because of decoherence ($Q=10^6$) before the verification pulses. \textbf{(b)} Generalization of the protocol, for $Q=10^6$ and $\chi_1=2$. The maximally reachable entanglement using a two-pulse entangling sequence is computed, for various measurement strength mismatches and resonator frequencies ratios. Maximal entanglement is set by the lowest measurement strength. Many configurations exist to generate entanglement between mechanical resonator, extending the validity of the protocol discussed in this paper to a more wider range of experimental situations. \label{chiImp}}
\end{figure}

In this section we investigate the effects of a mismatch between the measurement strengths for the two mechanical modes, which we model using the more general Hamiltonian
\begin{equation}
    H(t)=\frac{\hbar}{\sqrt{2}}  X_L\left(t\right)\sum_{j}g_j\left(X_j\cos{\left(\omega_jt\right)}+P_j\sin\left(\omega_jt\right)\right).
    \label{HamiltonianAsym}
\end{equation}

This can be taken into account in by applying an extra change of basis to write the mechanical eigenmodes of the Hamiltonian. Let's write $g = (g_1+g_2)/2$ and $g'=(g_1-g_2)/2$. The Hamiltonian then writes $H(t)=\hbar X_L\left(g X(t)+g' Y(t)\right)/\sqrt{2}$.
Then, any measurement that was planned to focus on $X_+$ for instance will always be associated with measurement of $X_-$ as well, with backaction on both $P_+$ and $P_-$ quadratures. This tends to reduce the squeezing of the two modes. The eigenmodes of the Hamiltonian are defined using an extra transform breaking the symmetry between the two modes 
\begin{equation}
    T=\frac{1}{\sqrt{g^{2}+g'^{2}}}\left(\begin{array}{cccc}
g & 0 & g' & 0\\
0 & g & 0 & g'\\
-g' & 0 & g & 0\\
0 & -g' & 0 & g
\end{array}\right).
\end{equation}

Figure \ref{chiImp}a shows the influence of this deviation on the logarithmic negativity of the mechanical state generated and verified by the ($k=0$, $l=0$) protocol, for ideally two-octaves-detuned resonators. The protocol generates maximal entanglement when the measurement strengths are equal. In the relevant regime of low decoherence and large thermal occupation, the amount of entanglement scales with the measurement strength mismatch $\epsilon_\chi=\chi_2/\chi_1-1$ as
\begin{equation}
    E_N\simeq-\mathrm{log}_2\left(\frac{1}{2\chi_1^2}\right)+\epsilon_\chi-2\chi_1^4\epsilon_\chi^2.
\end{equation}
At first order, entanglement tends to be stronger because of the larger average measurement strength of the two resonators, displacing the maximal reachable entanglement to higher values of $\chi_2/\chi_1$. Nevertheless, the maximum of entanglement is reached within a $\epsilon_\chi$ range of $\pm\sqrt{\log_2\left(2\chi_1^2\right)/2\chi_1^4}$, about $\pm10-30$\,\% for moderate measurement strengths considered in Fig.\,\ref{chiImp}a.

Noting that the two-pulse entanglement protocol (0,0) is robust to experimental deviations, Fig.\,\ref{chiImp}b extends the idea of two-pulse protocol to mechanical resonators with any measurement strength and frequency detuning. For each two-mode configuration, we computed the maximally reachable entanglement by means of a two-pulse sequence, if resonator $\omega_1$ has a measurement strength $\chi_1=2$. We observe that the maximal entanglement is set by the lowest measurement strength of the two resonators, and that many configurations actually leads to efficient mechanical entanglement. By adjusting the verification sequence to address different frequency mismatches, these results show that such a two-pulse protocol can be implemented in many experimental configurations to generate mechanical entanglement.

\section{Conclusion}

We propose and analyze in this work the use of two-pulse conditional measurement for bipartite Gaussian entanglement of two mechanical resonators coupled to an optical cavity. We focused on a specific protocol where the two resonators are detuned by two octaves, and couple to the optical cavity with equal strength. This specific protocol allows generation and verification of entanglement in less than a mechanical period, and shows robustness to experimental deviations of the mechanical frequency mismatch or optomechanical coupling mismatch. We showed that the two-pulse protocol can be extended to wider range of systems, by carefully choosing the two-pulse entanglement sequence and the verification sequence. The criteria to generate entanglement in terms of measurement strength and mechanical quality factor is reachable with current state of the art optomechanical designs. Moreover, this type of protocol may be extended to higher order entanglement generation in a local network of resonators, exploiting for instance higher order mechanical modes.
\ack
The authors thank Juha Muhonen and Albert Schliesser for insightful discussions. This work is part of the research programme of the Netherlands Organisation for Scientific Research (NWO). The authors acknowledge support from NWO Vidi, Projectruimte, and Vrij Programma (Grant No. 680.92.18.04) grants, the European Research Council (ERC starting grant no. 759644-TOPP), the European Union’s Horizon 2020 research and innovation programme under grant agreement no. 732894 (FET Proactive HOT), the Engineering and Physical Sciences Research Council (EP/T031271/1), UK Research and Innovation (MR/S032924/1) and the Royal Society.
\section*{References}
\bibliography{references.bib} 

\begin{thebibliography}{10}
\bibitem{note}
  One can indeed approximate the collective mode basis to the eigenbasis of the
  covariance matrix{.} At the end of the entangling sequence of duration
  $\tau=\pi/2\omega_1$, $\bm\sigma_m^\mathrm{ent}$ writes in the collective
  mode basis as \begin{equation} P(\tau)^{-1}\bm\sigma_m^\mathrm{ent}P(\tau) =
  \left(\begin{array}{cccc} \alpha_{+} & 0 & \alpha_{-} & 0\\ 0 &
  \alpha_{+}+\beta & 0 & \alpha_{-}\\ \alpha_{-} & 0 & \alpha_{+}+\beta & 0\\ 0
  & \alpha_{-} & 0 & \alpha_{+} \end{array}\right){.} \end{equation} Here
  again, the origin of the non-diagonal terms is because of the unbalance of
  the initial thermal occupation of the two modes. Then, in case of low thermal
  occupations, the collective mode basis covariance matrix is diagonal{.}
  Moreover, in the high occupation limit, the transfer matrix to go from
  $P(\tau)^{-1}\bm\sigma_m^\mathrm{ent}P(\tau)$ to the eigen-decomposition of
  $\sigma_m^\mathrm{ent}$ tends to the identity matrix, i{.}e{.} the collective
  mode basis is the eigenbasis.
\end{thebibliography}

\end{document}